\documentclass{iau}
\usepackage{graphicx}

\title[MCMC Constraints on LF Evolution] 
{Robust Constraint of Luminosity Function Evolution through MCMC Sampling}

\author[Noah Kurinsky \& Anna Sajina]   
{Noah Kurinsky \and Anna Sajina}

\affiliation{Department of Physics and Astronomy, Tufts University, \\ 4 Colby Street,
Medford, MA 02155, USA \\ Email: {\tt Noah.Kurinsky@gmail.com} }

\pubyear{2014}
\volume{306} 
\pagerange{}
\jname{Statistical Challenges in 21st Century Cosmology}
\editors{A.F. Heavens, J.-L. Starck, A. Krone-Martins eds.}
\begin{document}

\maketitle

\begin{abstract}
We present a new galaxy survey simulation package, which combines the power of Markov Chain Monte Carlo (MCMC) sampling with a robust and adaptable model of galaxy evolution. The aim of this code is to aid in the characterization and study of new and existing galaxy surveys. In this paper we briefly describe the MCMC implementation and the survey simulation methodology and associated tools. A test case of this full suite was to constrain the evolution of the IR Luminosity Function (LF) based on the HerMES (\textit{Herschel} SPIRE) survey of the \textit{Spitzer} First Look Survey field. The initial results are consistent with previous studies, but our more general approach should be of wider benefit to the community. 

\keywords{galaxies: luminosity function, galaxies: evolution, galaxies: statistics, methods: analytical, techniques: photometric, infrared: galaxies}
\end{abstract}

\firstsection 
\section{Motivation}

In extragalactic astronomy, we typically conduct surveys which uncover large numbers of high redshift galaxies with sparse SED coverage. Such spectra are insufficient for redshift determination by SED fitting, and spectroscopic follow-up to ascertain the redshifts for all galaxies in a survey is impractical, especially for samples of predominantly optically-faint IR galaxies. Direct measurements of luminosity function evolution are therefore very difficult in this regime. We can use well motivated galaxy spectral colors, however, to characterize cumulative trends in SED and luminosity function evolution in an entire population. By pairing a redshift-evolving SED library and a redshift parameterized luminosity function, we can simulate surveys given knowledge of evolutionary model parameters and local measurements. By sampling the resulting parameter space in a robust manner (as in \cite[Marsden \etal\ 2011]{marsden}), one can reconstruct the luminosity function as it evolves with redshift by comparison of the simulated and observed color-color space trends. In this manner, we can not only constrain the luminosity function parameters given observations, but also use the constrained parameters to simulate a high-redshift survey, given characteristics of the instrument performing the survey. 

The development of statistically sound fitting methods to accurately constrain the model parameters and generate model populations complicates such an effort, and requires the astronomer have extensive knowledge of a very specific class of statistical and computational methods. We aim to develop methods which make this analysis routine and easily adaptable so that such studies may be more efficient, robust, and accessible to time sensitive studies of new and existing surveys. We are developing a general software package capable of simulating a wide range of surveys with a robust Markov Chain Monte Carlo (MCMC) sampling algorithm, taking into account various LF forms, SED models, and instrumental properties, including noise levels and intensity limits. Our program allows us to characterize a survey as well as highlight the successes and failures of the SED template and the luminosity function form used by the simulation, and their behavior at high redshift, in order to speak to the nature of the data as well as the strength and weakness of the chosen models. We envision this program functioning, in the construction of high-redshift luminosity functions, similarly to how CosmoMC  (\cite{CosmoMC}) and the methods within are applied to constraint of $\Lambda$CDM parameters.

\section{Current Package}

Our simulation approach is centered around the choice of parameterized luminosity function and SED template library. The SED templates, as well as all other data inputs, are stored in FITS files and are easily interchangeable, specified as input parameters to the main program. The data to which parameters are fitted is photometric data in at least three bands, with observations in all bands. For a given set of luminosity function parameters, we follow the following procedure for each redshift-luminosity bin, with luminosity binning determined by the templates and redshift binning defined as a user specified parameter:
\begin{itemize}
\item Determine $N(L,z|\mathbf{x})$, the number of galaxies expected in a given luminosity and redshift bin given the luminosity function form and $\mathbf{x}$, the given parameters
\item Redshift the SED corresponding to the given $(L,z|\mathbf{x})$ bin to the observed frame, and convolve with instrumental filters
\item For each simulated galaxy, add Gaussian noise to each of three generated fluxes with noise determined by the mean of measurement errors found in the observations.
\item Determine detectability of the galaxy given survey flux cuts, and add the source to number count calculations if detected.
\end{itemize}
This procedure across the luminosity-redshift space results in a simulated survey with band-matched fluxes, which is projected into the same color space as the observations. We perform a modified $\chi^2$ test comparing data to observation to determine the quality of the fit.

To constrain the luminosity function parameters, we employ a parallel-chain MCMC simulation approach similar to that employed by \cite{CosmoMC}, but with updated methods to reflect the more recent findings of \cite{dunkley}. We employ the common Metropolis sampling algorithm, and the Gelman and Rubin convergence test. The user specifies likely initial parameter values, which parameters to fit, as well as acceptable ranges, and the program uses these as initial conditions. During the burn-in phase, one chain begins at the initial position and the others uniformly distributed throughout the allowable space, and the chains are allowed to freely explore as the temperature is annealed to achieve an optimal acceptance rate. Once the burn-in is complete, the chains are reinitialized closer to the best position, and the simulation runs until the convergence criteria are met. As the simulation progresses, covariance between parameters is continually adjusted to maximize sampling efficiency. 

The current package is structured as a script controlled C++ executable, with the main simulation and computationally expensive aspects implemented in C++ and the plotting and user interaction implemented in IDL and python. All data, SED models, and filters are read in as FITS or text files, and we aim to have fully selective and robust options for both of the latter two at first public release, expected to be complete by the end of the year. The luminosity functional form is hard coded in the C++ library, however the computational aspect is contained in a single function which is easily user editable, and the code can easily be recompiled by the standard GNU autotools.

\begin{figure}
\centering
\includegraphics[clip=true, bb=0 0 0 0, trim= 5 0 0 3, width=\textwidth]{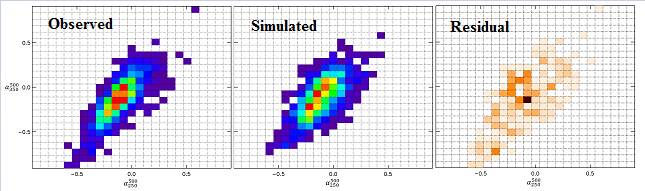}
\caption{Example of color-color histograms for an observed survey (left) and simulated survey (right), with the difference between the two shown at right. The free parameters are constrained by essentially attempting to minimize the cumulative sum of the right histogram, measured through our modified $\chi^2$ statistic. The discrepancy between these two metrics is on the order of $\sim$ 60 galaxies, compared to an observed sample of $\sim$ 1000 galaxies.}\label{fig1}
\end{figure}

\section{Initial Testing and Future Directions}
Our initial tests with a $\sim$1000 source HerMES (\textit{Herschel} SPIRE 250, 350, and 500 microns) survey, where only sources with measurements in all three bands were considered, were consistent with the density and luminosity evolution findings of \cite{marsden} as well as the SED intrinsic luminosity evolution found by \cite{sajina}. Using the locally derived SED templates of \cite{rieke}, we were able to reproduce the survey to within statistical margins. The observed and simulated survey metrics, along with their subtracted residual, can be seen in figure \ref{fig1}, where the total number of sources in the residual histogram is $\sim$ 60. We are continuing to refine this package, and aim to publish a paper in Fall 2014 detailing the first public release, and demonstrating test cases in the near and far infrared (\cite{kurinsky}). We are also working on testing the code on a wider range of surveys, including WISE (\cite{silva}).

\begin{acknowledgements}
We are grateful for the support of Tufts University's undergraduate research fund as well as the IAU Student Grant program for making this presentation possible. This work was funded in part by the Tufts Eliopoulos Summer Scholars grant and NSF grant 1313206.
\end{acknowledgements}

\end{document}